\documentclass{PoS}
\usepackage{amsmath}

\makeatletter
\def\simleq{\mathrel{\mathpalette\gl@align<}}
\def\simgeq{\mathrel{\mathpalette\gl@align>}}
\def\gl@align#1#2{\lower.6ex\vbox{\baselineskip\z@skip\lineskip\z@
     \ialign{$\m@th#1\hfill##\hfil$\crcr#2\crcr\sim\crcr}}}
\makeatother

\newcommand{\gf}{\gamma_5}

\newcommand{\nn}{\nonumber\\}

\newcommand{\bra}{\langle}
\newcommand{\ket}{\rangle}
\newcommand{\braket}[1]{\bra #1 \ket}

\newcommand{\xnuc}{\langle x \rangle}
\newcommand{\xxnuc}{\langle x^2 \rangle}

\newcommand{\xx}{\braket{x^2}}

\title{%
The calculation of nucleon strangeness form factors 
from $N_f=2+1$ clover fermion lattice QCD}

\ShortTitle{%
The calculation of nucleon strangeness form factors 
from $N_f=2+1$ clover fermion lattice QCD}

\author{\speaker{T. Doi}\\
Graduate School of Pure and Applied Science,
University of Tsukuba,
Tennodai 1-1-1, 
Tsukuba, Ibaraki 305-8571, Japan\\
        E-mail: \email{tdoi@het.ph.tsukuba.ac.jp}
}

\author{
{M. Deka},
{S.-J. Dong},
{T. Draper},
{K.-F. Liu}
and
{D. Mankame} \\
Department of Physics and Astronomy,
University of Kentucky, Lexington KY 40506, USA 
E-mail:
\email{mpdeka@pa.uky.edu},
\email{s.j.dong@uky.edu},
\email{draper@pa.uky.edu},
\email{liu@pa.uky.edu},
\email{dmmank2@uky.edu}
}

\author{N. Mathur\\
Department of Theoretical Physics,
Tata Institute of Fundamental Research,
Mumbai 40005, India\\
        E-mail: \email{nilmani@theory.tifr.res.in}
}

\author{T. Streuer\\
Institute for Theoretical Physics,
University of Regensburg, 93040 Regensburg, Germany\\
        E-mail: \email{thomas.streuer@physik.uni-regensburg.de}
}

\author{(${\rm \chi QCD}$ collaboration)}

\abstract{

We study the strangeness electromagnetic form factors
of the nucleon from the $N_f=2+1$ clover fermion lattice QCD calculation.
The disconnected insertions are evaluated using the Z(4) stochastic method,
along with unbiased subtractions from the hopping parameter expansion.
In addition to increasing the number of Z(4) noises,
we find that increasing the number of
nucleon sources for each configuration improves the signal significantly.
We obtain $G_M^s(0) = -0.017(25)(07)$, where the first error is statistical,
and the second is the uncertainties in $Q^2$ and chiral extrapolations.
This is consistent with experimental values,
and has an order of magnitude smaller error.
We also % present the % preliminary 
study the strangeness second moment of the partion distribution function
of the nucleon, $\xx_{s-\bar{s}}$.

}

\FullConference{The XXVII International Symposium on Lattice Field Theory - LAT2009\\
		 July 26-31 2009\\
		 Peking University, Beijing, China}

\begin{document}

\section{Introduction}

In the pursuit of the full understanding of QCD,
it has been essential to study the structure of the nucleon.
%and 
%experiments have been providing 
%many intriguing and unexpected results 
%for decades.
%
%
%
%[@
%For instance, 
%the EMC experiment~\cite{emc.89} marked the end of era
%when the nucleon could be explained as a mere constituent of three up and down quarks.
%%albeit the great success of the (naive) quark model.
%The consequence is that various QCD degrees of freedom in nucleon
%is important to achieve the full understanding.
%@]
%
%
%
The strangeness content of the nucleon 
particularly
attracts a great % deal of 
interest lately.
As the lightest non-valence quark structure,
it is an ideal probe
%to study 
%into 
for
the virtual sea quarks in the nucleon.
Recently, 
intensive experiments
have been carried out
for the electromagnetic form factors
%vector matrix elements, 
by
SAMPLE, %~\cite{sample.04}, % at MIT-Bates
A4, %~\cite{a4}, % at Mainz
HAPPEX, %~\cite{happex}, and
G0, %~\cite{g0.05}, % at JLab,
through parity-violating electron scattering (PVES).
The global analyses~\cite{young06,j.liu07,pate08} 
%have been performed
%and achieved a great accuracy, e.g.,
%have given, e.g.,
have produced, e.g.,
$G_E^s(Q^2) = -0.008(16)$ and
$G_M^s(Q^2) = 0.29(21)$ at $Q^2 = 0.1 {\rm GeV}^2$~\cite{j.liu07},
but substantial errors still exist so that the 
results are consistent with zero.
%
%
% SAMPLE(04):                G_M(s) [0.1      GeV^2] = +0.37  \pm 0.20 \pm 0.26 \pm 0.07 = 0.37(34)
% G0(05)    : G_E(s) + \eta  G_M(s) [0.12-1.0 GeV^2] = -0.05 - 0.10
% A4(04)    : G_E(s) + 0.225 G_M(s) [0.230    GeV^2] = +0.039 \pm 0.034
%             G_E(s)                [0.230    GeV^2] = +0.061 \pm 0.035 (if assuming G_M(s) is
%                            G_M(s) [0.230    GeV^2] = -0.099 \pm 0.154  by using HAPPEX G_E(s)+\eta G_M(s))
% A4(05)    : G_E(s) + 0.106 G_M(s) [0.108    GeV^2] = +0.071 \pm 0.036
%             G_E(s)                [0.108    GeV^2] = +0.032 \pm 0.051 (if using SAMPLE G_M(s))
% HAPPEX(06): G_E(s)                [0.091    GeV^2] = -0.038 \pm 0.042 \pm 0.010
%           : G_E(s) + 0.080 G_M(s) [0.099    GeV^2] = +0.030 \pm 0.025 \pm 0.006 \pm 0.012
%             G_E(s)                [0.1      GeV^2] = -0.01  \pm 0.03  (fit of various experiments)
%                            G_M(s) [0.1      GeV^2] = +0.55  \pm 0.28  (fit of various experiments)
% HAPPEX(07): G_E(s) + 0.09  G_M(s) [0.109    GeV^2] = +0.007 \pm 0.011 \pm 0.006
%             G_E(s)                [0.1      GeV^2] = -0.005 \pm 0.019
%                            G_M(s) [0.1      GeV^2] = +0.18  \pm 0.27
%
%
Making tighter constraints on these form factors from the theoretical side 
is one of the challenges in QCD calculation.
Moreover, such %theoretical 
constraints, together with experimental inputs,
can lead to more precise determinations of various interesting quantities,
such as the axial form factor $G_A^s$~\cite{pate08},
and the electroweak radiative corrections including 
the nucleon anapole moment, $\tilde{G}_A$~\cite{young06,PAVI02:musolf}.
Unfortunately, 
the theoretical status of strangeness form factors % , however,
remains quite uncertain. % to date.
For instance,
the values for the magnetic moment
$G_M^s(0)$ 
from different model calculations vary widely,
from $-0.5$ to $+0.1$.
%The analyses for the electric form factor are
%similarly ambiguous.
%
%
%
One may put an expectation on lattice simulation 
as the most desirable first-principle calculation.
The evaluation of strangeness matrix elements, however, has 
been a serious challenge to the lattice QCD,
since
it requires
the disconnected insertion (DI),
for which the straightforward calculation requires all-to-all propagators
and is prohibitively expensive.
%compared to the connected insertion (CI) calculation.
%
%
Consequently, there have been only few DI calculations, where
the all-to-all propagators are stochastically estimated,
and quenched approximation is used with Wilson fermion
~\cite{smm:ky_quenched,smm:ky_quenched2,smm:randy}.
There are also several indirect estimates
using quenched~\cite{smm:lein,wang08} or unquenched~\cite{huey07}
lattice data for the connected insertion (CI) part,
%with the help of 
and
the experimental magnetic moments (or electric charge radii) for octet baryons
as inputs. % under the assumption of isospin symmetry.
%These estimates obtained
%$G_M^s(0) = -0.046(19)$~\cite{smm:lein} and
%$G_M^s(0) = -0.066(26)$~\cite{huey07}.
%
%
%
In this proceeding,
we present 
the first full QCD lattice simulation of the direct DI calculation
for strangeness form factors. For details of this study, see Ref.~\cite{smm:doi}.

We also study the strangeness contribution to 
the lowest moments of the parton distribution function of the nucleon,
$\xnuc$, $\xxnuc$~\cite{x:doi}.
In particular, the strangeness second moment 
$\xx_{s-\bar{s}} = \int dx x^2 (s(x) - \bar{s}(x))$
is important quantity in relation to the NuTeV anomaly.
The NuTeV experiments~\cite{nutev} measured the Weinberg angle,
which is 3 $\sigma$ away from
%and obtained the value which is 3 $\sigma$ away from
the world average value.
While this may indicate the existence of New Physics,
it is necessary to make exhaustive investigations 
on hidden systematic uncertainties before making such a conclusion.
In fact, the asymmetry between strange and anti-strange parton distribution
is one of the probable candidates to
explain the NuTeV anomaly without New Physics.
We note that lattice QCD can make a great contribution to this issue,
by measuring the strangeness asymmetry in the second moment,
which provides essential constraint to the asymmetry in the strangeness 
parton distribution function.
In this proceeding, we present a preliminary result
of $\xx_{s-\bar{s}}$ as the first full QCD calculation 
for this quantity.

\section{Formalism and lattice calculation parameters}

We employ %the 
$N_f=2+1$ dynamical 
configurations
with nonperturbatively ${\cal O}(a)$ improved clover fermion 
and 
%renormalization group 
RG-improved gauge action
generated by CP-PACS/JLQCD Collaborations~\cite{conf:tsukuba2+1}.
We use %$(\beta,c_{sw})=(1.83,1.761)$
$\beta=1.83$ and
$c_{sw}=1.7610$ 
configurations 
with the lattice size of $L^3 \times T = 16^3\times 32$.
%In Ref.~\cite{conf:tsukuba2+1},
The lattice spacing was determined
as $a^{-1} = 1.625 {\rm GeV}$,
using 
$K$-input or
$\phi$-input~\cite{conf:tsukuba2+1}.
For the hopping parameters of $u$,$d$ quarks ($\kappa_{ud}$) and
$s$ quark ($\kappa_s$), we use
$\kappa_{ud} =$ $0.13825$, $0.13800$, and $0.13760$,
which correspond to $m_\pi =$ $0.60$, $0.70$, and $0.84$ ${\rm GeV}$,
respectively, and $\kappa_s = 0.13760$ is fixed.
We perform the calculation only at the dynamical quark mass points,
where
800 configurations are used for $\kappa_{ud} =$ $0.13760$,
and 810 configurations % are used
for
$\kappa_{ud} =$ $0.13800$, $0.13825$. % calculations.
The periodic boundary condition % (PBC)
is imposed
in all space-time directions
for the valence quarks.
%as was % imposed
%for the sea quarks.

%\begin{table}
%\caption{\label{tab:mass}
%The setup %calculation 
%parameters as well as % basic 
%hadron masses. % in lattice unit.
%$N_{noise}$ is the number of % Z(4) 
%noises in the stochastic estimate
%and $N_{src}$ is the number of %source points
%nucleon sources
%for each configuration.
%%in the two point function calculation.
%}
%%\begin{ruledtabular}
%\begin{tabular}{lcccccc}
%%$\kappa_{ud}$ & $N_{conf}$ & $N_{noise}$ & $N_{src}$ & $m_\pi a$ & $m_\rho a$ & $m_K a$ & $m_N a$ \\
%$\kappa_{ud}$ & $N_{conf}$ & $N_{noise}$ & $N_{src}$ & $m_\pi a$ & $m_K a$ & $m_N a$ \\
%\hline
%0.13760 & 800 & 600 & 64 & 0.5141(5) & 0.5141(5) & 1.0859(12) \\
%0.13800 & 810 & 600 & 82 & 0.4302(6) & 0.4540(5) & 0.9623(16) \\
%0.13825 & 810 & 800 & 82 & 0.3717(7) & 0.4141(6) & 0.8844(20) \\
%\end{tabular}
%%\end{ruledtabular}
%\vspace*{-3mm}
%\end{table}

We calculate the two point function (2pt) $\Pi^{\rm 2pt}$ 
and 
three point function (3pt) $\Pi^{\rm 3pt}_{J_\mu}$,
%and two point function (2pt) $\Pi^{\rm 2pt}$ defined as
%
%
%\begin{widetext}
%\vspace*{-2mm}
\begin{eqnarray}
\lefteqn{
\Pi^{\rm 3pt}_{J_\mu}(\vec{p},t_2;\ \vec{q},t_1;\ \vec{p'}=\vec{p}-\vec{q},t_0)
} \nn
&=& %\textstyle
\sum_{\vec{x_2},\vec{x_1}}
e^{-i\vec{p}\cdot(\vec{x}_2-\vec{x}_0)}
\times
e^{+i\vec{q}\cdot(\vec{x}_1-\vec{x}_0)}
\braket{0|{\rm T}\left[
\chi_N(\vec{x}_2,t_2) {J_\mu}(\vec{x}_1,t_1) \bar{\chi}_N(\vec{x}_0,t_0)
\right] |0}  ,
\label{eq:3pt}
%
%
%
%
%\nn[-7mm]
\end{eqnarray}
where 
%
%$\vec{p}, \vec{p}'$ and $\vec{q}$ are 
%the sink, source and insertion momentum, respectively,
%
%$\chi_N = \epsilon_{abc} (u_a^T C\gf d_b) u_c$ 
$\chi_N$ 
is the nucleon interpolation field % operator
and the insertion % current 
$J_\mu$ is given by the point-split conserved current
%
%\begin{eqnarray}
$
J_\mu (x+ \mu/2)
= 
(1/2) \times
$
{\small
$
\left[
\bar{s}(x+\mu) (1+\gamma_\mu) U_\mu^\dag(x) s(x)
- 
\bar{s}(x) (1-\gamma_\mu) U_\mu(x) s(x+\mu)
\right] .
\label{eq:vec}
$
}
%\end{eqnarray}
%

Electromagnetic form factors can be obtained 
%through the combination of 2pt and 3pt with 
using $\vec{p}=\vec{0}$, $\vec{p'}= -\vec{q}$ kinematics % (kinematic condition (A))
for the forward propagation ($t_2 \gg t_1 \gg t_0$)~\cite{smm:ky_quenched}.
In this work, we consider the backward propagation ($t_2 \ll t_1 \ll t_0$) as well,
in order to increase % the 
statistics. The formulas for 
%Sachs electric (magnetic) form factors
$G_E^s$ and $G_M^s$
%$G_E = F_1 - \frac{E^q_N-m_N}{2m_N} F_2$,
%$G_M = F_1 + F_2$,
are summarized as
%
%
%\vspace*{-2mm}
\begin{eqnarray}
%
%
%
%\lefteqn{
%
R_\mu^{\pm} (\Gamma_{\rm pol}^\pm) &\equiv&
\frac{
{\rm Tr}\left[
\Gamma_{\rm pol}^\pm\cdot
\Pi^{\rm 3pt}_{J_{\mu}}(\vec{0},{t}_2;\ \pm\vec{q},{t}_1;\ -\vec{q},t_0)
\right]
}
{
{\rm Tr} \left[
\Gamma_e^\pm\cdot \Pi^{\rm 2pt}_{}(\pm\vec{q},t_1;\ t_0)
\right]
}
%
%} 
%
%
%\cdot
%\qquad
\times
\frac{
{\rm Tr} \left[
\Gamma_e^\pm\cdot \Pi^{\rm 2pt}_{}(\vec{0},t_1;\ t_0)
\right]
}
{
{\rm Tr} \left[
\Gamma_e^\pm\cdot \Pi^{\rm 2pt}_{}(\vec{0},t_2;\ t_0)
\right]
} ,
\label{eq:ratio,kin1,fb} \\
%
%
%
%\nn[-8mm]
%\end{eqnarray}
%
%\vspace*{-5mm}
%
%\begin{eqnarray}
%
%
R_{\mu=4}^{\pm}(\Gamma_{\rm pol}^\pm=\Gamma_e^\pm) &=& \pm G_E^s(Q^2), 
\qquad
R_{\mu=i}^{\pm}(\Gamma_{\rm pol}^\pm=\Gamma_k^\pm) =
\frac{\mp \epsilon_{ijk} q_j}{E^q_N+m_N}
G_M^s(Q^2) ,
\label{eq:ele1,mag,kin1,fb} %\\ %[-1mm]
%
%
%R_{\mu=i}^{\pm}(\Gamma_{\rm pol}^\pm=\Gamma_k^\pm) &=& 
%\frac{\mp \epsilon_{ijk} q_j}{E^q_N+m_N}
%G_M^s(Q^2) ,
%\label{eq:mag,kin1,fb}
%
%
%\nn[-8mm]
\end{eqnarray}
where 
$\{i,j,k\} \neq 4$,
$
\Gamma_e^\pm \equiv (1\pm\gamma_4)/2
$ ,
$
\Gamma_k^\pm \equiv (\pm i)/2 \times (1\pm\gamma_4) \gf \gamma_k
$ 
and
%
%$m_N$ denotes the nucleon mass, 
$E^q_N \equiv \sqrt{m_N^2 + \vec{q}^{\,2}}$.
%
%
%
%For the choice of double sign,
%The forward propagation corresponds to the upper sign
%and the backward % propagation corresponds 
%to the lower sign.
The upper (lower) sign corresponds to 
the forward (backward) propagation.
Furthermore, we consider 
another kinematics of
$\vec{p}=\vec{q}$, $\vec{p'}=\vec{0}$,
% (kinematic condition (B)).
%
%
%
where the analogs of 
Eqs.~(\ref{eq:ratio,kin1,fb}),
(\ref{eq:ele1,mag,kin1,fb})
hold.
We find that the results from 
the latter kinematics
%Eq.~(\ref{eq:ratio,kin1,fb})
%kinematical condition (B)
%Eqs.~(\ref{eq:ele1,kin2,fb}), (\ref{eq:mag,kin2,fb})
have similar size of statistical errors as those from
the former,
%Eq.~(\ref{eq:ratio,kin2,fb}),
%condition (A),
%Eqs.~(\ref{eq:ele1,kin1,fb}), (\ref{eq:mag,kin1,fb}),
and the average of them yields better results.
Hereafter, we present results from total average of 
two kinematics and forward/backward propagations,
unless otherwise noted.
%
%Therefore, compared to the previous studies in the literature,
%our results have about factor 4 enhancement in statistics.
%
%
%
%
%
In order to extract the matrix elements, % from Eq.~(\ref{eq:ele1,mag,kin1,fb})
%with various $t_1, t_2$ results,
we 
take the 
summation over the insertion time $t_1$,
%which 
symbolically 
given as
%written as
%
%\begin{eqnarray}
%\begin{equation}
$
R^{t}_{E,M} \equiv
\frac{1}{K_{E,M}^\pm}
\sum_{t_1=t_0+t_{s}}^{t_2-t_{s}}
R^{\pm}_\mu
=
{\rm const.} + t_2 \times G_{E,M}^s ,
\label{eq:tsum}
$
%\end{equation}
%\end{eqnarray}
%
where $K_{E,M}^\pm$ are trivial kinematic factors 
%which appear in
appearing in
Eq.~(\ref{eq:ele1,mag,kin1,fb}) %and (\ref{eq:mag,kin1,fb}),
and $t_s$ is chosen so that the error is minimal.
%From Eq.~(\ref{eq:tsum}), 
%From the above equation,
We thus obtain $G_{E,M}^s$ as 
the linear slope of $R^{t}_{E,M}$ against $t_2$.
%In order to achieve ground state saturation,
%we use the data only for $t_2 \geq 7$~\cite{x:deka}.
%~\cite{x:deka}.

%
\begin{figure}[bt]
\begin{center}
\includegraphics[width=0.3\textwidth,angle=270]{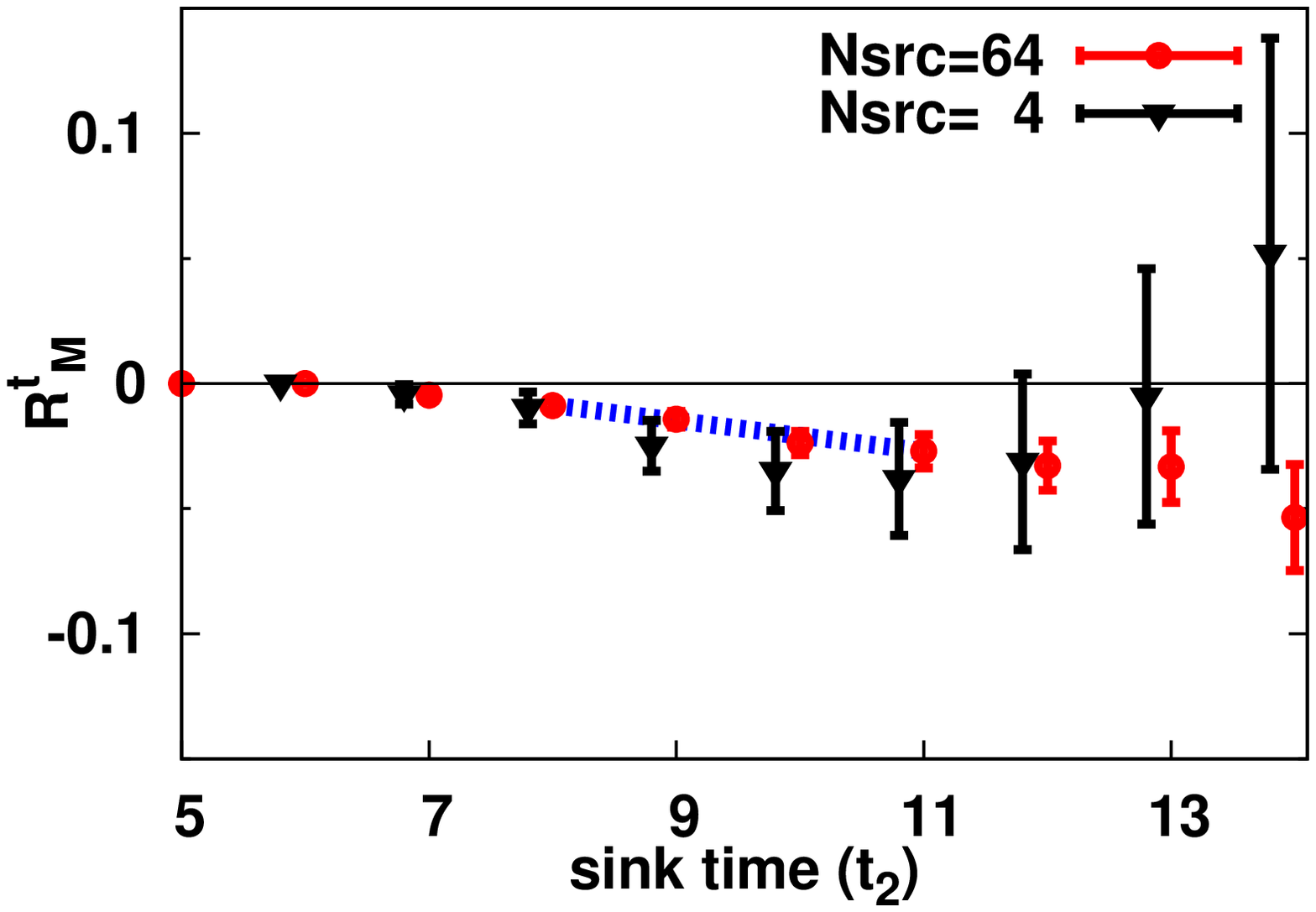}
%\hspace*{-1em}
\includegraphics[width=0.3\textwidth,angle=270]{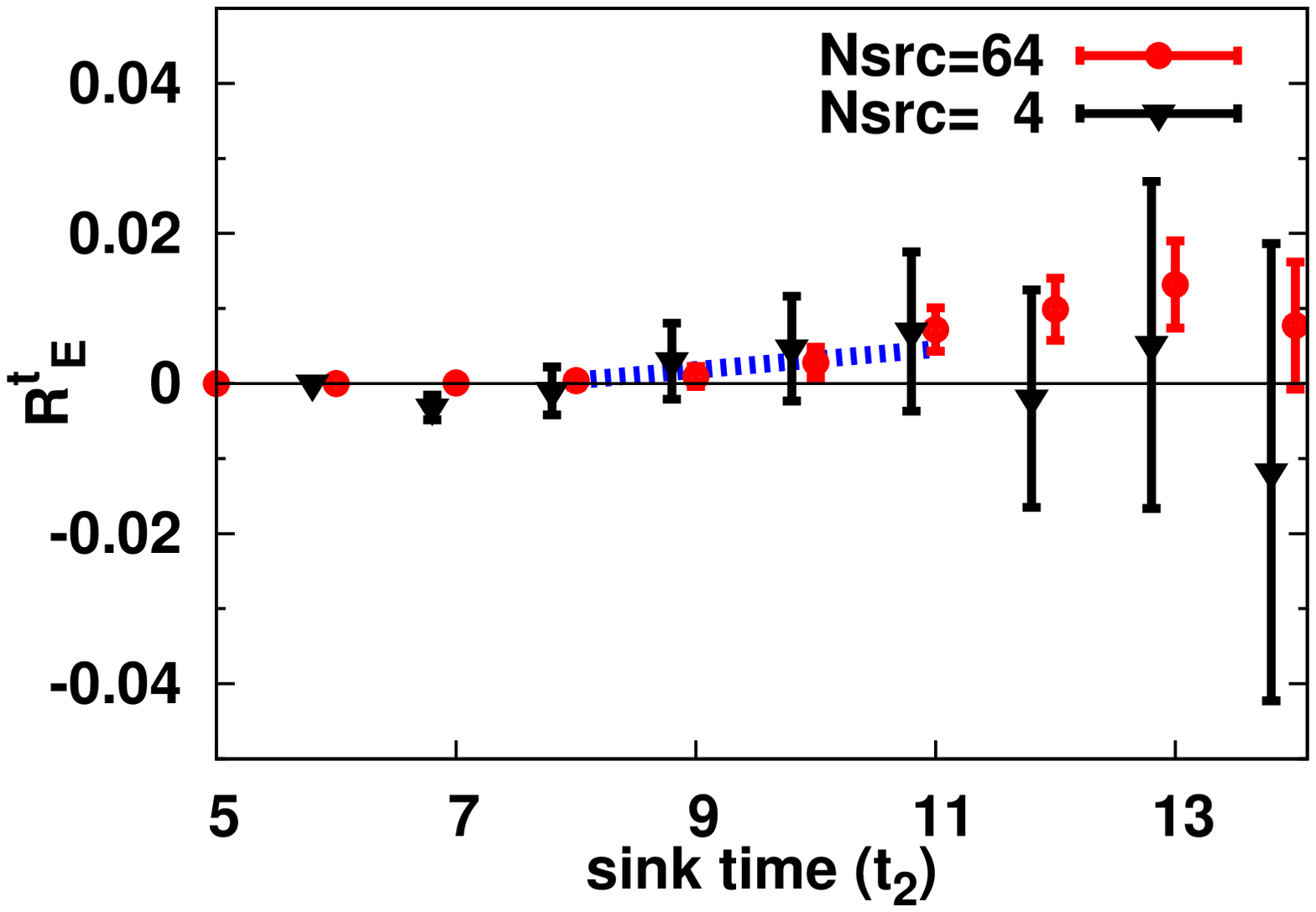}
\end{center}
\vspace*{-8mm}
\caption{
 \label{fig:tsum}
%Lattice data for 
$R_M^t$ (left) and $R_E^t$ (right) 
%as obtained from Eq.~(\ref{eq:tsum}) 
%in Eq.~(\ref{eq:tsum}) 
with %$(\kappa_{ud},\kappa_s) = (0.13760, 0.13760)$,
$\kappa_{ud}=0.13760$, 
%$n=2$, 
$\vec{q}^{\,2}=2\cdot (2\pi/La)^2$,
$N_{src}=64$ (circles) and $N_{src}=4$ 
(triangles, with offset for visibility),
plotted against the nucleon sink time $t_2$.
The dashed line is the linear fit
where the slope corresponds to the form factor.
%The results from $N_{src}=4$ are also plotted 
%(triangles, with offset for visibility), 
%in order to show the effectiveness of increasing $N_{src}$.
\vspace*{-3mm}
}
\end{figure}

The calculations of 3pt 
%for the strangeness current
need
the evaluation of DI.
We use the stochastic method~\cite{DI:noise},
with Z(4) noises in color, spin and space-time indices.
We generate independent noises for different configurations,
in order to avoid possible auto-correlation.
%We use $N_{noise} = 600$ noises for $\kappa_{ud} = 0.13760, 0.13800$
%and $N_{noise}=800$ for $\kappa_{ud} = 0.13825$ calculations.
%
%
To reduce fluctuations, 
we use the charge conjugation and $\gamma_5$-hermiticity (CH), and parity 
symmetry~\cite{smm:doi,angular:ky_quenched,x:deka}.
%For instance, we find that the information for the $G_M^s$ is coded
%in the product of ${\rm Re}(\Pi^{\rm 2pt}) \times {\rm Re}({\rm loop})$,
%and filtering out the imaginary parts reduces the noises~\cite{angular:ky_quenched,x:deka}.
%$
%\Pi^{\rm 3pt}_{J_\mu} 
%= \Pi^{\rm 2pt} \times ({\rm loop})
%= ({\rm Re}(\Pi^{\rm 2pt}) + i {\rm Im}(\Pi^{\rm 2pt})) 
%\times ({\rm Re}({\rm loop}) + i {\rm Im}({\rm loop}))
%$,
%and identify the necessary one term for the form factor of concern,
%with the help of the charge conjugation and $\gamma_5$-hermiticity (CH) symmetry 
%and parity symmetry. 
%In this way, we can eliminate the unnecessary noise % contamination 
%from the other three terms~\cite{x:deka}.
%For details, see Refs.~\cite{DI:noise}.
%
%
We also perform unbiased subtractions~\cite{DI:hpe} to 
%offset 
reduce
the off-diagonal contaminations to the variance.
For subtraction operators,
we employ those obtained %through
from
hopping parameter expansion (HPE) for the propagator $M^{-1}$, %~\cite{DI:hpe},
%
%\begin{eqnarray}
$
\frac{1}{2\kappa} M^{-1} = \frac{1}{1+C} + \frac{1}{1+C} (\kappa D) \frac{1}{1+C} 
+ \cdots
%+  \frac{1}{1+C} (\kappa D) \frac{1}{1+C} (\kappa D) \frac{1}{1+C} + \cdots
%
$
%\end{eqnarray}
%
where $D$ denotes the Wilson-Dirac operator and $C$ the clover term.
We subtract up to order $(\kappa D)^4$ term,
and % observe that 
the statistical error is 
reduced
%suppressed 
%by as much as factor 2.
by a factor of 2.

%
%
%
%\begin{table*}[tbh]
%\caption{\label{tab:s_em}
%Lattice results for the strangeness electromagnetic form factors
%with the momentum-squared $\vec{q}^{\,2} = n\cdot ( 2\pi/ La)^2$.
%}
%\begin{tabular}{cccccccccc}
%
%                & \multicolumn{4}{c}{$G_M^s(Q^2) (\times 10^{-2})$} & \multicolumn{5}{c}{$G_E^s(Q^2) (\times 10^{-2})$} \\
%$\kappa_{ud} \backslash n$ & 1 & 2 & 3 & 4                   & 0 & 1 & 2 & 3 & 4 \\ \hline
%0.13760 & -0.96(29) & -0.61(17) & -0.57(20) & -0.20(25)      &  0.21(21)  &  0.01(10) & 0.14(08) & 0.22(11) & 0.10(15) \\
%0.13800 & -0.76(37) & -0.76(24) & -0.57(32) &  0.04(41)      &  0.15(28)  &  0.16(14) & 0.14(12) & 0.15(19) & 0.46(29) \\
%0.13825 & -1.09(41) & -1.02(27) & -0.67(33) & -0.25(47)      & -0.04(33)  & -0.16(15) & 0.36(13) & 0.27(20) & 0.71(31)
%
%\end{tabular}
%\vspace*{-3mm}
%\end{table*}
%
%
%
%
%
%
%
%
%\begin{table}
%\vspace*{-1mm}
%\caption{\label{tab:Q2fit}
%The parameters fitted against $Q^2$ behavior. % of $G_{E,M}^s(Q^2)$.
%}
%\begin{tabular}{ccccccc}
%              & \multicolumn{3}{c}{$G_M^s(Q^2)$}              & \multicolumn{2}{c}{$G_E^s(Q^2)$} \\
%$\kappa_{ud}$ & $G_M^s(0)$         & $\Lambda a$ & $\chi^2/{\rm dof}$ & $g_E^s$            & $\chi^2/{\rm dof}$ \\
%              & $(\times 10^{-2})$ &             &                    & $(\times 10^{-2})$ &                    \\
%\hline
%0.13760 & -1.7(12) & 0.66(29) & 0.34(83)      & 1.2(5) & 0.56(89) \\
%0.13800 & -1.4(09) & 0.77(40) & 0.77(126)     & 1.4(6) & 0.34(70) \\
%0.13825 & -1.9(11) & 0.80(40) & 0.38(87)      & 1.9(7) & 2.68(189)
%\end{tabular}
%\vspace*{-3mm}
%\end{table}
%
%
%

In the stochastic method, 
it is quite expensive
to achieve a good signal to noise ratio (S/N) just by increasing $N_{noise}$
because S/N improves with $\sqrt{N_{noise}}$.
In view of this, we use many nucleon point sources $N_{src}$ 
in the evaluation of the 2pt part for each configuration~\cite{x:deka}.
Since the calculations of the loop part and 2pt part
are independent of each other, this is expected to be an efficient way. % to increase statistics.
In particular, for the $N_{noise} \gg N_{src}$ case,
we observe that S/N improves almost ideally, by a factor of $\sqrt{N_{src}}$.
We take $N_{src}=64$ for $\kappa_{ud} = 0.13760$
and $N_{src}=82$ for $\kappa_{ud} = 0.13800, 0.13825$, % calculations,
where locations of sources are taken 
so that they are 
% dispersed 
separated
in 4D-volume as much as possible.
%
%
%
%The % lattice 
%calculation parameters 
%as well as basic hadron masses are tabulated in 
%Tab.~\ref{tab:mass}.

%There are several ways~\cite{smm:ky_quenched,smm:randy} to extract the matrix elements from 
%Eq.~(\ref{eq:ele1,mag,kin1,fb}) % and (\ref{eq:mag,kin1,fb})
%with various $t_1, t_2$ results.
%Among them, it is advocated~\cite{smm:ky_quenched,smm:ky_quenched2,angular:ky_quenched,x:deka} to 

%%%%%%%%%%%%%% Results %%%%%%%%%%%%%%%%%%

\section{Results for strangeness form factors}

We calculate 
for the 
five smallest momentum-squared points,
%
%\begin{eqnarray}
$
\vec{q}^{\,2} = n\cdot ( 2\pi/ La)^2\ %\qquad 
$
%$(n=0,1,2,3,4)$ , 
%(n\in \{0,4\}),
$(n=0$--$4)$.
%\end{eqnarray}
%
%which roughly corresponds to 
%$Q^2 = 0, 0.8, 1.1, 1.4{\rm GeV^2}$.
%
%which correspond to 
%$|\vec{q}| = 0, 0.64, 0.90, 1.1, 1.3$ GeV.
%
% kap=0.13760: Q^2 = 0.3947, 0.7672, 1.1208, 1.4582 GeV^2
% kap=0.13800: Q^2 = 0.3916, 0.7560, 1.0983, 1.4221 GeV^2
% kap=0.13825: Q^2 = 0.3889, 0.7469, 1.0804, 1.3938 GeV^2
%
%
Typical figures for $R^{t}_{E,M}$ % $R^{t}_{E,M}$ 
are shown in Fig.~\ref{fig:tsum}.
One can 
observe 
the significant S/N improvement by increasing $N_{src}$.
% is also observed.
%
%The numerical results are given
%in Tab.~\ref{tab:s_em}.
%
%We find
%$G_E^s(0)$
%are consistent with zero, 
%which serves as a test of the calculations.
%
Of particular interest is that,
for all $\kappa_{ud}$ simulations,
$G_M^s(Q^2)$ is found 
to be negative with 
%two to three
2-3 
$\sigma$ signals for low $Q^2$ regions.
%
%On the other hand, $G_E^s(Q^2)$ is found to be quite small in general.
%
%
%
%
%
In order to determine the magnetic moment, % $G_M^s(0)$, 
the $Q^2$ dependence of $G_M^s(Q^2)$ is studied.
We employ the dipole form in the $Q^2$ fit,
$G_M^s(Q^2) = G_M^s(0) / (1+Q^2/\Lambda^2)^2$,
where reasonable agreement with lattice data is observed.
%where correlations among 
%different $Q^2$ are taken into account.
%
%
%
For the electric form factor,
we employ 
$G_E^s(Q^2) = g_E^s \cdot Q^2 / (1+Q^2/\Lambda^2)^2$,
considering that $G_E^s(0)=0$ from the vector current 
conservation.
In the practical fit of $G_E^s(Q^2)$, however, 
reliable extraction of the pole mass $\Lambda$ is impossible
because $G_E^s(Q^2)$ data are almost zero within error.
Therefore, we assume that $G_E^s(Q^2)$ %the electric form factor
has the same pole mass as $G_M^s(Q^2)$, %the magnetic form factor,
and perform a one-parameter fit for $g_E^s$~\cite{smm:doi}.
%The obtained parameters are given in Tab.~\ref{tab:Q2fit}.
%We also test the simultaneous fit of 
%$G_M^s(Q^2)$ and $G_E^s(Q^2)$ with 
%three parameters of $G_M^s(0), \Lambda, g_E^s$, 
%obtain the consistent results with
%those from the separate fit.

Finally, we perform the chiral extrapolation for the fitted parameters.
Since our quark masses are relatively heavy, 
we consider only the leading dependence on $m_K$,
which is obtained by heavy baryon chiral perturbation theory (HB$\chi$PT). %~\cite{chPT:hemmert}.
For the magnetic moment $G_M^s(0)$, we fit linearly in terms of $m_K$~\cite{smm:ky_quenched,chPT:hemmert}.
For the pole mass $\Lambda$, we take that the magnetic mean-square radius
%$\braket{r^2_s}_M \equiv -6 \frac{d G_M^s}{d Q^2}|_{Q^2=0} = 12 G_M^s(0) / \Lambda^2$
$\braket{r^2_s}_M = 12 G_M^s(0) / \Lambda^2$
behaves as $1/m_K$~\cite{chPT:hemmert}.
For $g_E^s$, we use the electric radius
%$\braket{r^2_s}_E \equiv -6 \frac{d G_E^s}{d Q^2}|_{Q^2=0} = -6 g_E^s$
$\braket{r^2_s}_E = -6 g_E^s$
which has an $\ln (m_K/\mu)$ behavior~\cite{chPT:hemmert}, and we take the scale $\mu=1$ GeV.
The chiral extrapolated results are
$G_M^s(0) = -0.017(25)$,
$\Lambda a = 0.58(16)$,
$\braket{r_s^2}_M = -7.4(71) \times 10^{-3} {\rm fm}^2$,
%and
%$g_E^s = 0.027(16)$.
$\braket{r_s^2}_E = -2.4(15) \times 10^{-3} {\rm fm}^2$.

%%%%%%%%%%%%%%%%%%%%%%%%%%%%%%%%%%%%%%%%%%%%%%%%
%\section{Discussion}

%In Ref.\cite{leinweber}, the 
%relation between strangeness electromagnetic form factors
%and the ratio of DI(u,d) to DI(s) are given
%using the isospin symmetry and experimental inputs
%for magnetic moment.
%
%Our results for DI(u,d) / DI(s) are XXX,
%which is consistent/inconsistent from the input
%in Ref.\cite{leinweber}.
%This can explain why our results are different/same
%as those from Ref.\cite{leinweber}.
%
%Also, our DI(u,d) / DI(s) indicates that
%the seq quark contribution to the magnetic moment
%is rather small/large,
%because there is cancellation
%as $\mu(sea) = +2/3 G_M^{u,DI} - 1/3 G_M^{d,DI} - 1/3 G_M^s$.
%This may explain why even naive quark model can reproduce
%the baryon magnetic moment relatively within good approximation (???).

%
%
%
\begin{figure}[tb]
\begin{center}
%
%\vspace*{-8mm}
%\hspace*{-10mm}
\includegraphics[width=0.28\textwidth,angle=270]{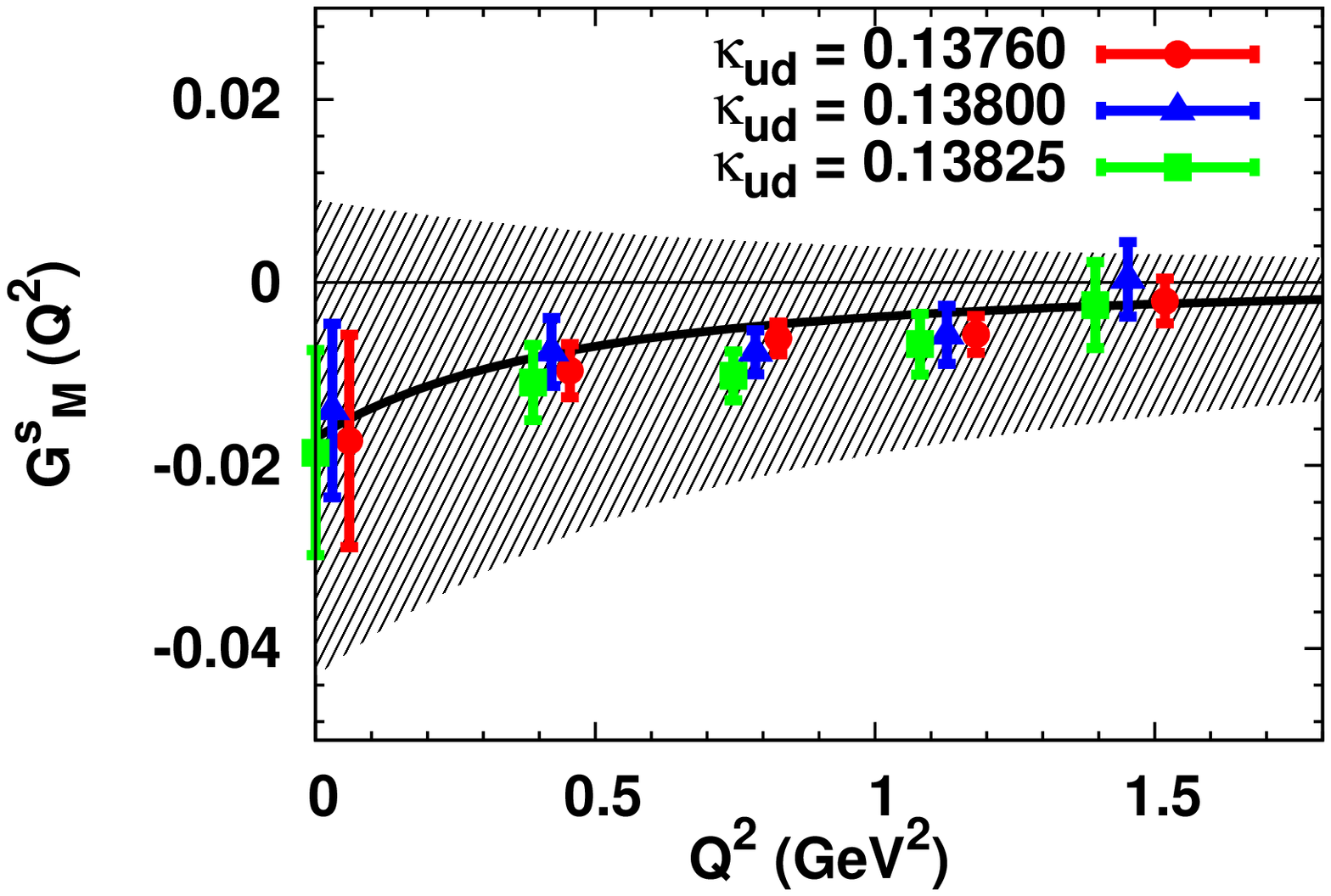} %\\[-3mm]
%\vspace*{-5mm}
%\hspace*{-10mm}
\includegraphics[width=0.28\textwidth,angle=270]{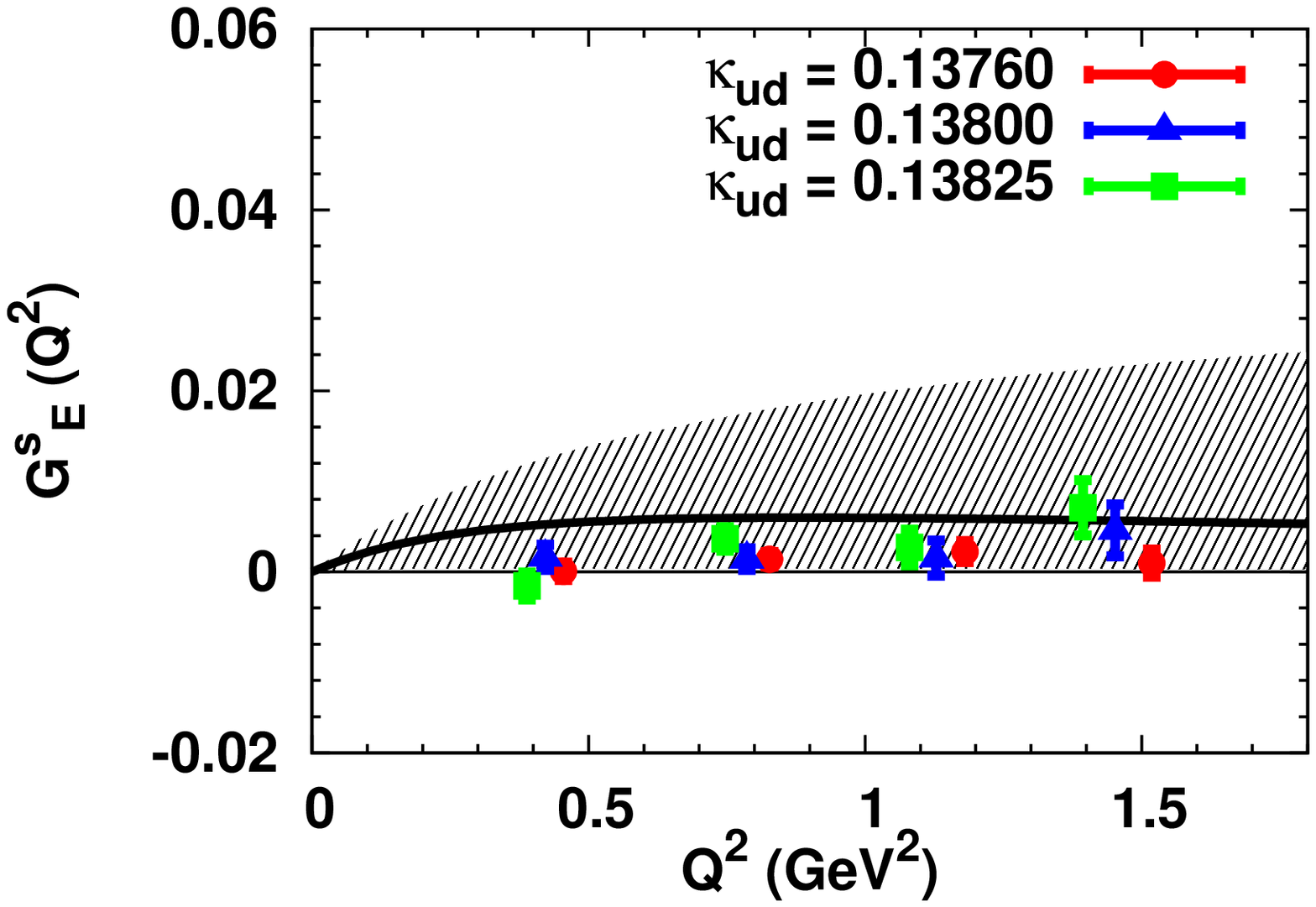}
\vspace*{-5mm}
\end{center}
\caption{
 \label{fig:res}
The chiral extrapolated results for $G_M^s(Q^2)$ (left) 
and $G_E^s(Q^2)$ (right) plotted with solid lines.
Shaded regions 
represent
%the one-$\sigma$ uncertainty 
the error-band 
with statistical and systematic error
added in quadrature.
%Plotted 
Shown
together are the lattice data 
(and $Q^2$-extrapolated $G_M^s(0)$)
for
$\kappa_{ud}=$ 
0.13760 (circles),
0.13800 (triangles),
0.13825 (squares)
with offset for visibility.
%For the $G_M^s(Q^2)$ figure, the $Q^2$ extrapolated 
%data for each $\kappa_{ud}=$ are also shown.
\vspace*{-3mm}
}
\end{figure}

Before quoting the final results,
we consider the systematic 
uncertainties yet to be addressed.
First, we analyze the ambiguity of $Q^2$ dependence in form factors,
by %which is rather unknown for the strangeness.
employing the monopole form~\cite{smm:doi}.
%$G_M^s(Q^2) = G_M^s(0) / (1+Q^2/\tilde{\Lambda}^2)$,
%$G_E^s(Q^2) = g_E^s\cdot Q^2 / (1+Q^2/\tilde{\Lambda}^2)$,
%and find that the precision of the lattice data cannot 
%disentangle the difference of the dipole/monopole behaviors.
The results are %from the monopole fit are
found to be consistent with those from the dipole fit,
and we take the difference as 
systematic uncertainties.
Second, we study the uncertainties in chiral extrapolation
by testing two alternative extrapolations.
In the first one,
we take into account 
the nucleon mass dependence on the quark mass,
using the lattice nucleon mass. 
From the physical viewpoint,
this corresponds to measuring the magnetic moment 
not in units of 
lattice %nuclear 
magneton but physical
magneton~\cite{qcdsf:formfactor}.
In the second alternative,
we use the linear fit %wrt. 
in terms of 
$m_K^2$, %\cite{rbc2+1:formfactor},
observing the results have weak quark mass dependence.
In either of alternative analyses, we find that
the results are consistent with previous ones.
While a further clarification with physically light quark mass simulation
and a check on convergence of HB$\chi$PT~\cite{chPT:hammer}
is desirable, 
%given that the current study is performed in relatively
%heavy quark mass region, 
we use the dependence of results on different extrapolations
as systematic uncertainties. % in the chiral extrapolation.
Third, we examine the contamination from excited states.
Because our spectroscopy study % for 2pt
indicates that the mass of Roper resonance is
massive compared to the $S_{11}$ state
on the current lattice~\cite{ky:meng},
the dominant contaminations are (transition) form factors 
associated with $S_{11}$.
%$N \to S_{11}, S_{11} \to N, S_{11} \to S_{11}$.
%
%
On this point, we find that 
such contaminations can be eliminated theoretically,
making the appropriate substitutions for 
$\Gamma_e^\pm$ in Eq.~(\ref{eq:ratio,kin1,fb})
and
$\{ \Gamma_e^\pm,\ \Gamma_k^\pm \}$ in Eq.~(\ref{eq:ele1,mag,kin1,fb})~\cite{smm:doi}.
It is found that 
the results from this formulation are basically the same as before,
so we conclude that the contamination regarding the $S_{11}$ state is 
negligible.

%Fourth, we consider the dependence of results
%on the $s$ quark mass, because the simulation point 
%is somewhat lighter than the physical 
%$s$ quark mass point~\cite{conf:tsukuba2+1,conf:tsukuba2+1:pre}.
%We first note that the $s$ quark loop is expected
%to be relatively suppressed for a larger $s$ quark mass. In fact, 
%a separate calculation with different valence $s$ quark mass
%yields consistent results with this argument.
%Therefore, $G_{E,M}^s$ at physical $s$ mass
%correspond to smaller values in absolute magnitude,
%and such effect will be automatically 
%subsumed
%in the final error analysis
%since the results given below (almost) contain $|G_{E,M}^s|=0$ in the error-bands.
%(i.e., the results always fall in the shaded region in Fig.~\ref{fig:res}.)

As remaining sources of systematic error,
one might worry that 
the finite volume artifact could be substantial considering that 
the spacial size of the lattice is about $(2 {\rm fm})^3$.
%On this point, 
However,
we recall that 
Sachs radii are found to be quite small, 
$ |\braket{r_s^2}_{E,M}| \ll 0.1 {\rm fm}^2$,
%which is self-consistent with the small finite volume artifact.
which %might 
indicates a small finite volume artifact.
For the discretization error,
we conclude that finite $(qa)$ discretization error is negligible,
since the lattice nucleon energy % at each $\vec{q}$ on the lattice
is found to be consistent with the dispersion relation.
As another discretization error,
we note that $m_N$ ($m_K$) is found to have 6 (8) \% error for the current 
configurations~\cite{conf:tsukuba2+1,conf:tsukuba2+1:pre}. 
Considering the dependence of $G_{E,M}^s$ on these masses,
we estimate that the discretization errors % in our results 
amount to $\simleq 10$\%, and are much smaller than the statistical errors.
Of course, more quantitative investigations 
%on these issues 
are desirable,
%are necessary,
%with larger and finer lattices,
and such work is in progress.

To summarize the results of form factors,
we obtain 
%For the magnetic moment, 
$G_M^s(0) = -0.017(25)(07)$, 
where the first error is statistical and
the second is systematic %error 
from uncertainties of the $Q^2$ extrapolation and chiral extrapolation.
We also obtain $\Lambda a = 0.58(16)(19)$ for dipole mass
or $\tilde{\Lambda} a = 0.34(17)(11)$ for monopole mass,
and $g_E^s = 0.027(16)(08)$.
These lead to 
%from which we obtain 
$G_M^s(Q^2) = -0.015(23)$,
$G_E^s(Q^2) =  0.0022(19)$ at $Q^2=0.1 {\rm GeV}^2$,
where error is obtained by quadrature from 
statistical and systematic errors.
%where the error is the squared sum of 
%the statistical and
%systematic error.
%
%
We also obtained, e.g.,
$G_M^s(Q^2) = -0.014(21)$,
$G_E^s(Q^2) =  0.0041(38)$ at $Q^2=0.22 {\rm GeV}^2$.
Note that 
these are consistent with the world averaged data 
at $Q^2=0.1 {\rm GeV}^2$~\cite{young06,j.liu07,pate08}
and the recent measurement at Mainz~\cite{a4:new},
$G_M^s(Q^2) = -0.14(11)(11)$,
$G_E^s(Q^2) =  0.050(38)(19)$ at $Q^2=0.22 {\rm GeV}^2$,
with an order of magnitude smaller error.
In Fig.~\ref{fig:res}, 
we plot our results for
$G_M^s(Q^2)$, $G_E^s(Q^2)$,
where the shaded regions 
correspond to the square-summed error.

\section{Results for the strangeness second moment $\xx_{s-\bar{s}}$}

In the calculation of the strangeness second moment, $\xx_{s-\bar{s}}$,
we use the following three-index operator $T_{4ii}$ as
the insertion operator in 3pt,
\begin{eqnarray}
T_{4ii} &\equiv&
-\frac{1}{3}
\left[ 
  \bar{s} \gamma_4   \overleftrightarrow{D}_i   \overleftrightarrow{D}_i   s
+ \bar{s} \gamma_i   \overleftrightarrow{D}_4   \overleftrightarrow{D}_i   s
+ \bar{s} \gamma_i   \overleftrightarrow{D}_i   \overleftrightarrow{D}_4   s
\right] \qquad (i\neq 4).
\end{eqnarray}
Taking the kinematics of $\vec{p'} = \vec{p}$, $\vec{q}=\vec{0}$,
the following ratio of 3pt to 2pt corresponds to the second moment~\cite{x:deka},
\begin{eqnarray}
\frac{
{\rm Tr}\left[
\Gamma_e^\pm\cdot
\Pi^{\rm 3pt}_{T_{4ii}}(\pm\vec{p},t_2;\ \vec{0},t_1;\ \pm\vec{p},t_0)
\right]
}
{
{\rm Tr} \left[
\Gamma_e^\pm\cdot \Pi^{\rm 2pt}(\pm\vec{p},t_2;\ t_0)
\right]
}
=
\pm p_i^2 \cdot \xx_{s-\bar{s}} ,
\end{eqnarray}
where the upper (lower) sign corresponds to 
the forward (backward) propagation as before.

In the evaluation of the DI, 
we can apply basically the same technique which is used 
in the calculation of the strange form factors.
Because of the difference of the structure of the insertion operator,
we perform the unbiased subtraction from HPE 
up to order $(\kappa D)^3$ term.

\begin{figure}[bt]
\begin{center}
\includegraphics[width=0.3\textwidth,angle=270]{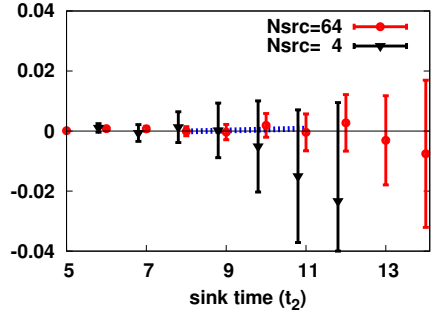}
%\hspace*{-1em}
%
\end{center}
%\vspace*{-8mm}
\caption{
 \label{fig:tsum:x2}
The ratio of 3pt to 2pt 
with 
$\kappa_{ud}=0.13760$, 
$\vec{p}^{\,2}=(2\pi/La)^2$,
$N_{src}=64$ (circles) and $N_{src}=4$ 
(triangles, with offset for visibility),
plotted against the nucleon sink time $t_2$.
The dashed line is the linear fit
where the slope corresponds to the 
second moment.
}
\end{figure}

In Fig.~\ref{fig:tsum:x2} , we plot the ratio of 3pt to 2pt
in terms of the nucleon sink time, $t_2$, where
the linear slope corresponds to the signal of $\xx_{s-\bar{s}}$.
One can clearly observe that 
S/N improves significantly by increasing $N_{src}$,
as was observed in the form factor study.
Yet, the signal of $\xx_{s-\bar{s}}$ in this figure is still consistent with zero.
In order to obtain the final result,
detailed analyses are in progress.
%consideration of renormalization and 
%the chiral extrapolation is in progress.

%

\section{Summary}

We have studied the strangeness electromagnetic form factors
of the nucleon from the $N_f=2+1$ clover fermion lattice QCD calculation.
It has been found that calculating many nucleon sources is essential to achieve 
a good S/N in the evaluation of DI.
We have obtained the form factors 
which are consistent with experimental values,
and have an order of magnitude smaller error.
The importance of the strangeness second moment, $\xx_{s-\bar{s}}$,
has been emphasized, 
and a preliminary result has been reported.

\begin{acknowledgments}
We thank the CP-PACS/JLQCD Collaborations
for their configurations.
This work was supported  in part by 
U.S. DOE grant DE-FG05-84ER40154
and Grant-in-Aid for JSPS Fellows 21$\cdot$5985. % (11315615).
Research of N.M. is supported by Ramanujan Fellowship.
The calculation was performed 
%on supercomputers 
at Jefferson Lab, Fermilab
and 
%the University of Kentucky,
the Univ. of Kentucky,
partly using the Chroma Library~\cite{chroma}.
\end{acknowledgments}

\end{document}